\documentclass[conference]{IEEEtran}
\IEEEoverridecommandlockouts
\usepackage{cite}
\usepackage{amsmath,amssymb,amsfonts}
\usepackage{algorithmic}
\usepackage{graphicx}
\usepackage{textcomp}
\usepackage{xcolor}
\usepackage{algorithm}

\usepackage{amsmath}
\usepackage{listings}
\usepackage{xcolor}
\usepackage{amsthm}
\usepackage{multirow}
\usepackage{diagbox}
\usepackage{xfrac}
\usepackage{caption,tabularx,booktabs}
\usepackage{soul}

\definecolor{codegreen}{rgb}{0,0.6,0}
\definecolor{codegray}{rgb}{0.5,0.5,0.5}
\definecolor{codepurple}{rgb}{0.58,0,0.82}
\definecolor{backcolour}{rgb}{0.95,0.95,0.92}

\definecolor{variableColor}{rgb}{0.56,0.10,0.70}

\theoremstyle{definition}
\newtheorem{definition}{Condition}[section]

\lstdefinestyle{mystyle}{
    commentstyle=\color{codegreen},
    keywordstyle=\color{magenta},
    numberstyle=\tiny\color{codegray},
    stringstyle=\color{codepurple},
    basicstyle=\ttfamily\footnotesize,
    breakatwhitespace=false,
    breaklines=true,
    captionpos=b,
    keepspaces=true,
    numbers=left,
    numbersep=5pt,
    showspaces=false,
    showstringspaces=false,
    showtabs=false,
    tabsize=2
}

\newfloat{listing}{!ht}{lop}
\floatname{listing}{Listing}

\def\BibTeX{{\rm B\kern-.05em{\sc i\kern-.025em b}\kern-.08em
    T\kern-.1667em\lower.7ex\hbox{E}\kern-.125emX}}
\begin{document}

\title{Vectorization Of Narrow Matrix Multiplication for Ascend AI Inference Acceleration}

\author{\IEEEauthorblockN{Shurygin Anton}
\IEEEauthorblockA{
\textit{Moscow Institute of Physics and Technology}\\
Moscow, Russia \\
shurygin.aa@phystech.edu}
\and
\IEEEauthorblockN{Frolov Aleksandr}
\IEEEauthorblockA{
\textit{Moscow Institute of Physics and Technology}\\
Moscow, Russia \\
aksdr.frolov@phystech.edu}}

\maketitle

\begin{abstract}
This research proposes and evaluates a novel approach to optimizing matrix multiplication (MatMul) on Huawei Ascend NPUs, motivated by a key insight:
during matrix-vector multiplication (narrow MatMul), the Cube Unit (AIC) is often underutilized, while the Vector Unit (AIV) remains idle for most of the operator runtime.
In this paper, we introduce the MatMul algorithm, which uses vector instructions of AscendC to effectively offload computations from the Cube Unit to the Vector Unit.
The algorithm was tested and applied to accelerating the inference of MLA DeepSeek-V3 operator.
By successfully overlapping AIV and AIC computations, our optimization showed a mean performance gain of 20\% for a single token processing scenario.
Our work addresses a significant gap in the literature on practical optimization techniques for AscendC, despite the availability of documentation and the active CANN community.
\end{abstract}


\section{Introduction}
Optimizing matrix multiplication (MatMul), a fundamental and resource-intensive operation,
is crucial for the efficiency of neural networks, especially large language models (LLMs)\@.
To address this problem, specialized hardware accelerators have emerged, in particular neural processing units (NPUs~\cite{tan2021efficient}) and tensor processing units (TPUs~\cite{jouppi2017datacenter}), which use dedicated cores specifically designed to accelerate linear algebra operations.
While extensive optimization strategies that leverage warp-level primitives and tensor cores exist for GPUs~\cite{kerr2017cutlass},
and TPUs exploit systolic arrays for massive throughput~\cite{jouppi2017datacenter},
the NPU architecture requires different approaches for breakthrough optimizations~\cite{huang2025towards},~\cite{moustafa2023accelerating}.

Huawei Ascend NPU~\cite{liao2021ascend} series is one of the leading platforms in this area.
During our work on optimizing operators for Ascend hardware, we made a key observation based on operator profiling on an architectural simulator.
When a matrix is multiplied by a vector (either a row or a column), the efficiency of the powerful hardware unit responsible for matrix calculations, known as Cube, is significantly reduced.
At the same time, while the operator is executing, the Vector hardware unit often remains idle.
This observation led us to hypothesize that significant performance gain could be achieved by offloading such MatMul to the vector executor.

In this paper, we propose a solution that eliminates the problem of extremely low utilization of the Cube Unit (AIC) during matrix-vector multiplication (i.e.\ narrow MatMul).
We implemented and optimized the MatMul algorithm for the Vector Unit (AIV) using the vector instructions available in the AscendC API\@.
The key optimization factor is that by offloading the narrow MatMul operation from AIC to AIV, subsequent independent matrix calculations can be started much earlier on the Cube.
Thus, by successfully overlapping the computations on the Cube Unit and Vector Unit, it will be possible to achieve significant acceleration of operator inference.

To the best of our knowledge, there are very few similar studies or practical implementations that use low-level vector optimization techniques for accelerating the MatMul operation on the AscendC.
Moreover, this deficit remains despite the fact that there is open documentation on AscendC with samples and best practices~\cite{cannsamples}.
Therefore, this paper also expands the scope of potential related studies that could be benefit for both industry and academia.

We summarize our contributions as follows:
\begin{itemize}
\item We discovered a novel approach to optimizing matrix multiplication for mathematical operators implemented using the AscendC API\@.
\item We designed a generalized MatMul algorithm for AIV, that expands the space of possible optimizations in the CANN community~\cite{canncommunity}.
\item Narrow MatMul optimization was applied to accelerating the MLA DeepSeek-V3 operator inference and showed a mean performance gain of 20\% for a single token processing scenario.
\end{itemize}

\section{Background}

\subsection{Hardware Architecture}

\begin{figure}[htbp]
\centerline{\includegraphics[width=0.5\textwidth]{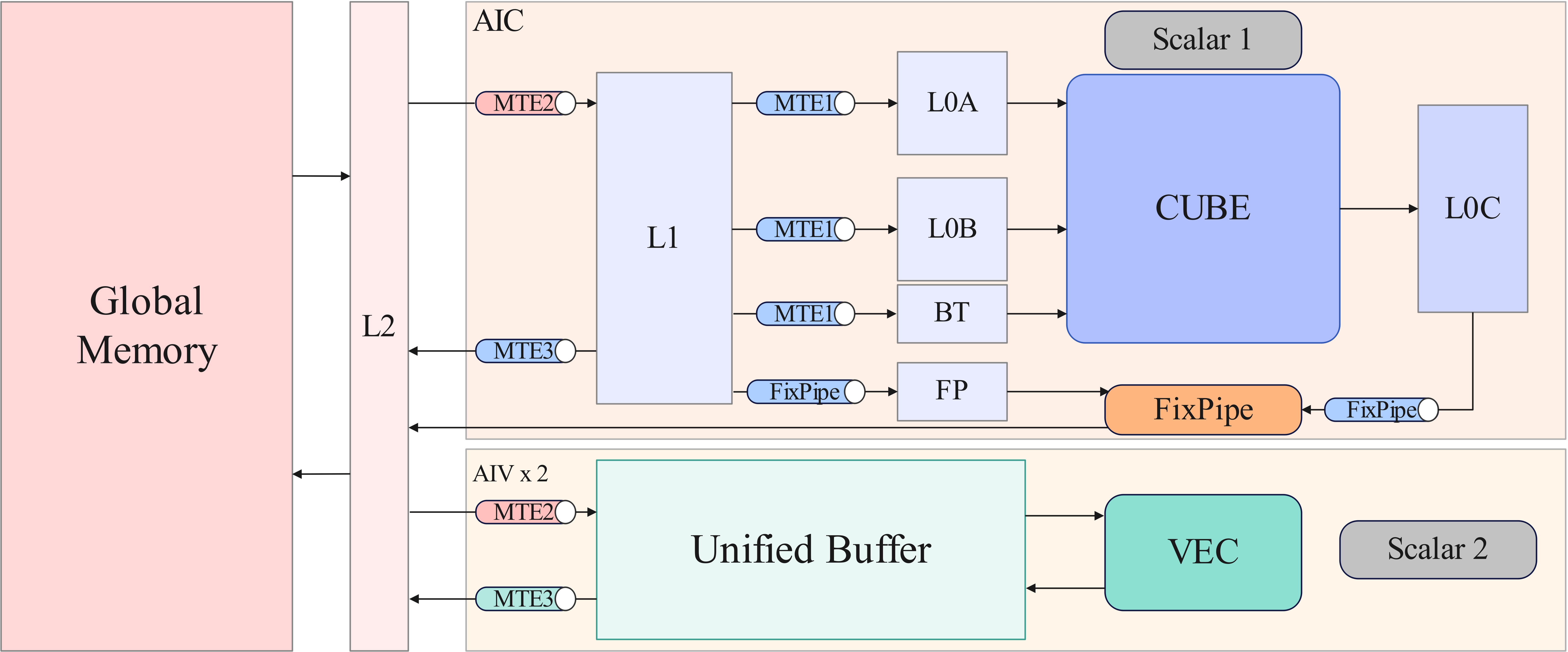}}
\caption{AI core architecture (Ascend 910B)}\label{HardwareArch}
\end{figure}

The Ascend AI processor is a System on Chip~\cite{liang2020ascend},~\cite{liao2021ascend}.
Its main architectural components include special computing units, large-capacity storage units, and the corresponding control units.
The processor can be divided into Control CPU, AI Computing Engine (including AI Core and AI CPU),
multilevel on-chip system cache, etc.
AI Core adopts the DaVinci architecture, whose basic structure is shown in Fig.~\ref{HardwareArch}.
It includes three basic computing resources: Cube Unit, Vector Unit, and Scalar Unit.

The AIC and AIV are the main computational units of the AI Core.
The Cube Unit mainly performs matrix-related operations, the Vector Unit is responsible for vector operations (e.g., calculating trigonometric functions),
and Scalar Unit handles all types of scalar data operations and program control flow.

\subsection{Cube Unit}\label{subsec:cube_overview}

The Cube Unit provides powerful parallel multiplications and additions, enabling AI Core to finish matrix computations rapidly.
Cube is able to achieve throughput of one $16\times 16$ multiplication per cycle~\cite{liang2020ascend}.
At the time of this research, the AIC on the modern Ascend 910B chip processor supports a wide range of data types, including INT4, INT8, FP16, BF16, FP32, etc.

\subsection{Vector Unit}\label{subsec:vector_overview}
The Vector Unit in AI Core is mainly responsible for performing vector-related operations.
Each Ascend chip contains two independent Vector Units (AIV$\times$2), each with its own Unified Buffer (UB) with a size of 192 kilobytes.
In one instruction, a Vector Unit can process no more than 256 consecutive bits.
However, the Vector Unit's instruction set has limitations; for instance, basic arithmetic operations are not supported for the BF16 type.

\subsection{Matrix Tiling}

Due to the limited capacity of the on-chip cache, the constraints of the computing and memory resources on the processor, it is often necessary to split the matrix into tiles.
The purpose of this method is to fully exploit the principle of data locality 
by reusing the matrix block data and subsequently accelerating access through caching.
In addition to this, the tiling abstraction is crucial for optimally distributing computations across AI cores.

\begin{figure}[htbp]
    \centerline{\includegraphics[width=0.5\textwidth]{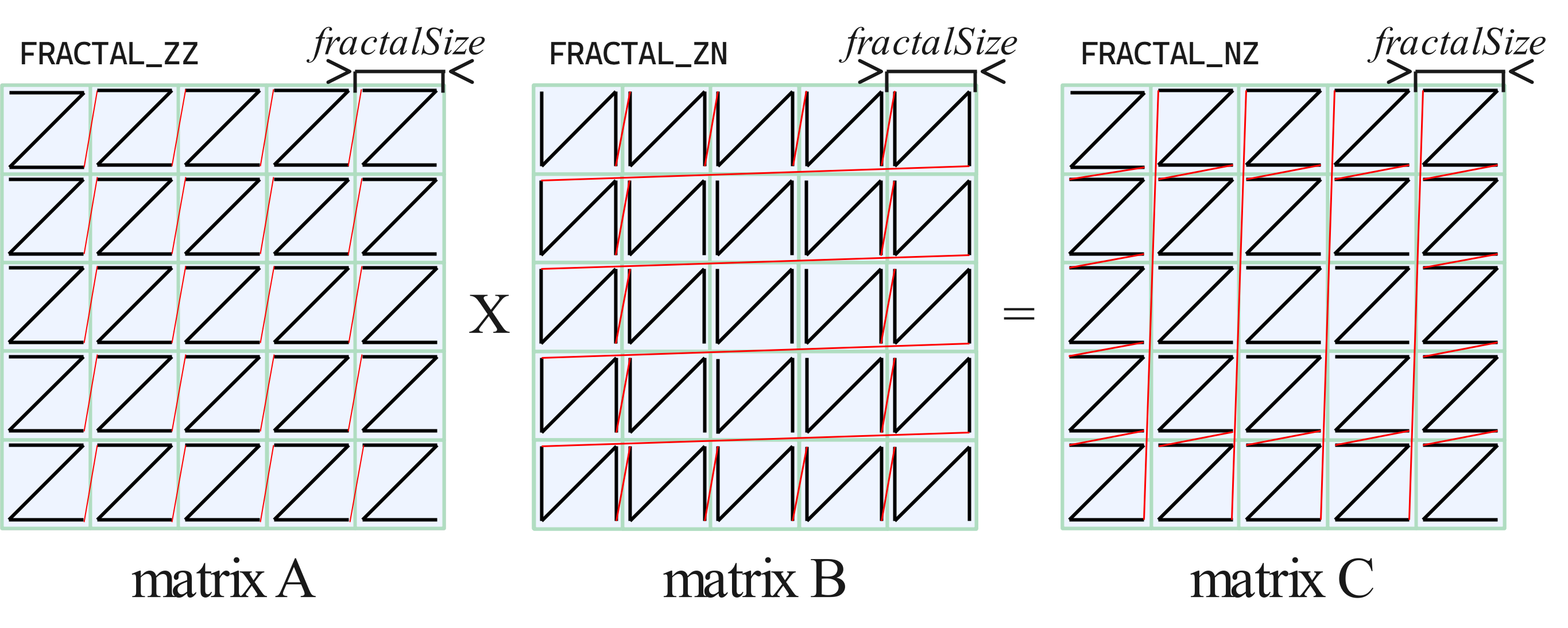}}
    \caption{Special formats related to matrix multiplication}\label{MatrixDataLayout}
\end{figure}

When performing MatMul using the basic AscendC API, there are certain requirements for the data format of the input and output tensors.
In our work we consider the tiling of two-dimensional tensors.
Fig.~\ref{MatrixDataLayout} shows one possible matrices partitioning when multiplying $A \times B = C$ on the Cube Unit.
\begin{itemize}
    \item The partitioning of the matrix $A$ according to the AscendC data layout format must be \texttt{FRACTAL\textunderscore{}ZZ}.
    \item Each block of matrix $B$ is sorted by rows, while the inner part of each block is sorted by columns, which is known \texttt{FRACTAL\textunderscore{}ZN}.
    \item The resulting matrix $C$ is arranged as each block matrix being partitioned by columns, and the data inside each block being partitioned by row, so-called \texttt{FRACTAL\textunderscore{}NZ}.
\end{itemize}

The open source code~\cite{cannopsadvrepo} shows specific applications of these formats in developing related operators using the AscendC API\@.

\subsection{Problem Statement}
When splitting matrices, it is possible that the tiling parameters will not be multiples of the matrix size.
In this case, missing rows or columns are padded with zeros to ensure that the Cube Unit works correctly.
Consequently, for matrix-vector multiplication, the computational efficiency of the Cube Unit drops dramatically by 16 times, as it effectively processes only one vector of a $16\times 16$ block, leaving $\frac{15}{16}$ of its capacity idle.

Considering the diversity of modern hardware with different amounts of AI Cores, optimization issues often involve tilings where Cube Unit performance suffers due to small remnants and excessive padding.
Therefore, we come to the need to consider alternative approaches of computing narrow MatMul.
And since AIVs are one of the main forces of Ascend AI, the optimization through offloading narrow MatMul from Cube Unit to Vector Unit looks promising.

\section{Vector MatMul Optimization}

As evident from the architectural analysis, a naive offloading of MatMul from AIC to AIV is not beneficial.
According to a rough estimate, a single Vector Unit is approximately 64 times slower than a single Cube Unit.
However, MatMul is a core operation in modern LLMs, especially within complex operators, such as Attention mechanism~\cite{vaswani2017attention},~\cite{dao2022flashattention}.\@
In general, we can conclude that the optimization is worth it if MatMul calculation on AIV \textbf{can overlap with subsequent operator computations} on AIC\@.
The formalized necessary optimization conditions for successful offloading of MatMul from AIC to AIV are stated in the following Section~\ref{subsec:opt_scope}.

\subsection{Optimization Scope}\label{subsec:opt_scope}

Optimization by offloading MatMul $A \times B = C$ from AIC to AIV is worth it if the following conditions are met simultaneously.

\begin{definition}[Data Independence]\label{condition1}
{
Subsequent resource-intensive \textbf{operator computations} on the Cube Unit must not depend on the result of the calculation of the vectorized MatMul.
}\end{definition}

\begin{definition}[Narrow Dimension]\label{condition2}
{
Let $\mathit{A} \in \mathbb{R}^{M \times K}$, $\mathit{B} \in \mathbb{R}^{K \times N}$.
Then the condition is considered satisfied if one of the matrices is a vector, i.e., $\min(M, N) = 1$.
}\end{definition}

\begin{definition}[Idling]\label{condition3}
{
If $P$ physical AI cores are used to execute an operator, then by definition the hardware has $2P$ Vector Units (\ref{subsec:vector_overview}).
Let $V_{idle}$ be the number of idle vectors during the execution of the operator.
Then the condition is considered satisfied if $V_{idle} \approx 2P$.
}\end{definition}

\subsection{Proposed Algorithm}\label{sec:proposed_algo}

\begin{figure}[htbp]
    \centerline{\includegraphics[width=0.5\textwidth]{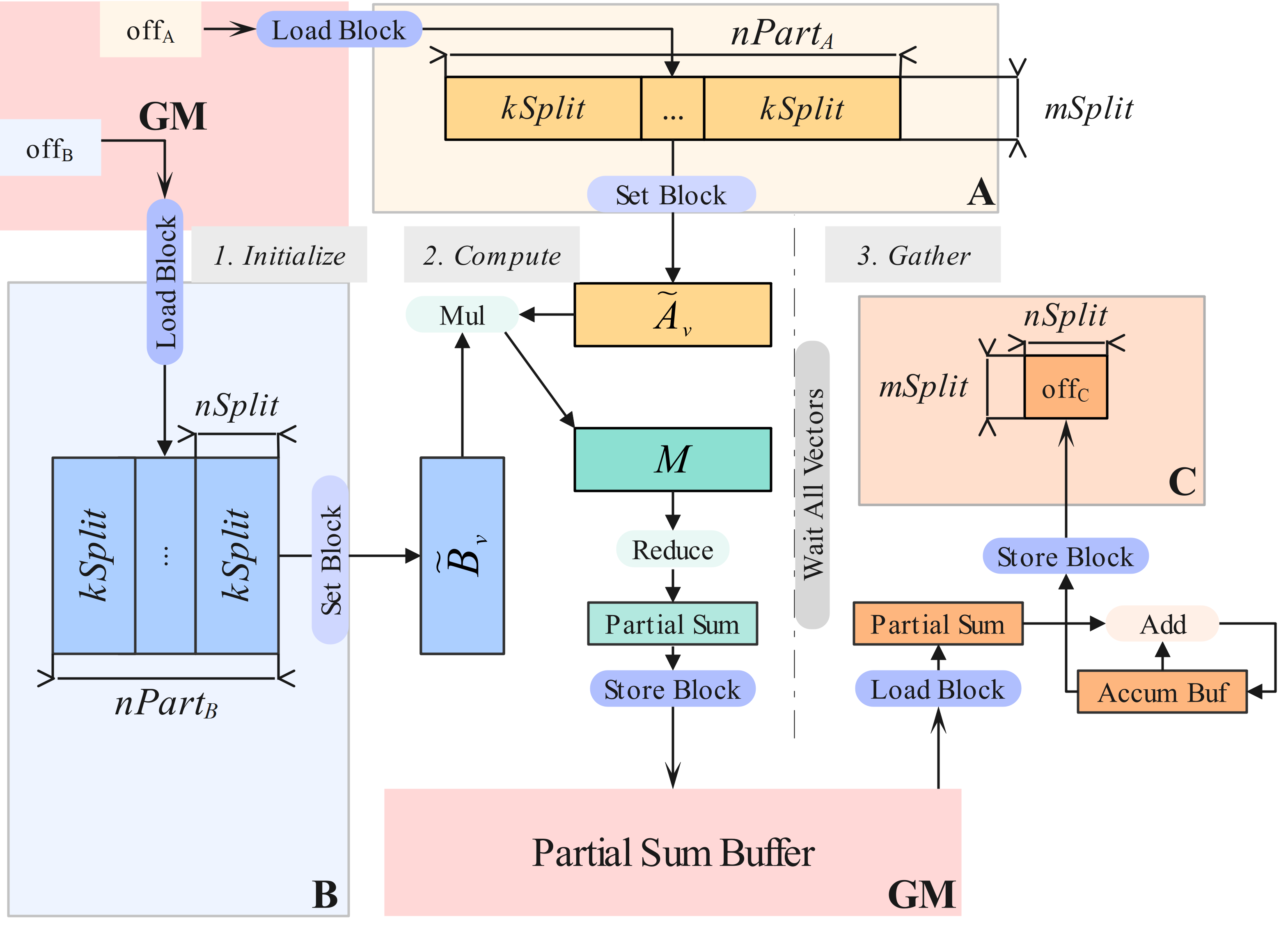}}
    \caption{Generalized MatMul algorithm scheme on AIV}\label{MatMulScheme}
\end{figure}

The scheme of the generalized matrix multiplication algorithm is shown in Fig.~\ref{MatMulScheme}.
Logically, the algorithm can be divided into three stages: initialization, computation and gather.
All these stages are described in the pseudocode of the generalized solution given in Algorithm~\ref{alg:generalized_algo}.
It starts with the \textbf{initialization stage}.
Depending on the AIV identifier of the executor ($v$), it is necessary to load the blocks of matrices $A$ and $B$ from Global Memory (GM) into the Unified Buffer at the corresponding offsets ($off_{A}$, $off_{B}$ respectively).

After loading all the input data, the algorithm moves to the \textbf{computation stage}.
Since the Vector Unit can process no more than $256$ consecutive bits in one instruction (sec.~\ref{subsec:vector_overview}), it is necessary to split the loaded block of matrices $A_{v}$, $B_{v}$, respectively.
Depending on the matrices tiling and the number of available AI cores ($P$), parameters must be chosen that determine how many calculations each AIV should do.
Thus, in general case, at the computation stage, each Vector Unit perform at least $nPartA \cdot nPartB$:
\begin{itemize}
    \item vector multiplications of rows ($\widetilde{A_{v}}$) by columns ($\widetilde{B_{v}}$)
    \item reductions of the result of multiplication ($M$) to a vector of partial sums ($partSum$)
    \item stores of partial sums into a temporary buffer in Global Memory ($tmpBuf$)
\end{itemize}

At the end of the Algorithm~\ref{alg:generalized_algo}, the data \textbf{gathering stage} is implemented.
After finishing the multiplication of matrix blocks, all Vector Units must wait for each other.
Once synchronized, a sufficient number of Vector Units begin to gather the columns of the resulting matrix $C$ block-by-block, accumulating partial sums loaded from the Partial Sum Buffer ($tmpBuf$).

\section{DeepSeek-V3 Inference Optimization}\label{optimization_application}

The practical relevance of this research is demonstrated by its successful application on Ascend NPU hardware to solve a performance tuning problem.
The application of the developed optimization allowed us to accelerating the inference~\cite{li2024large} of the DeepSeek-V3 for an important edge case, processing of a single token.

\begin{algorithm}[H]
\caption{Generalized MatMul on AIV}\label{alg:generalized_algo}
\begin{algorithmic}[1]
\REQUIRE $\mathit{A} \in \mathbb{R}^{M \times K}$, $\mathit{B} \in \mathbb{R}^{K \times N}$, $\{k,m,n\}Split$
\COMMENT{$\{k,m,n\}Split$ \text{-- partitioning parameters}}

\ENSURE $\mathit{C} \in \mathbb{R}^{M \times N}$

\textit{Init stage}

$\mathit{P}$ \text{-- number of AI cores}

$\mathit{v}$ \text{-- AIV identifier, so} $0 \leq \mathit{v} \leq {2P - 1}$

\text{Calculates the offset in GM for block assigned to $\mathit{v}$}
\STATE $\mathit{off_A} \gets \textsc{getOff}_A(\mathit{v})$
\STATE $\mathit{off_B} \gets \textsc{getOff}_B(\mathit{v})$

\text{Loads tensor block from GM by offset}
\STATE $\mathit{A_{v}} \gets \textsc{LoadBlock}(A, off_A)$
\STATE $\mathit{B_{v}} \gets \textsc{LoadBlock}(B, off_B)$

\textit{Compute stage}

\FOR {$\mathit{bIdx} = 0$ \TO $\mathit{nPart_{B}} - 1$}
    \STATE $\widetilde{B_{v}} \gets \textsc{SetBlock}(\mathit{B_{v}}, bIdx)$
    \FOR {$\mathit{aIdx} = 0$ \TO $\mathit{nPart_{A}} - 1$}
        \STATE $\widetilde{A_{v}} \gets \textsc{SetBlock}(\mathit{A_{v}}, aIdx)$

        \STATE $\mathit{M} \gets \textsc{Mul}(\widetilde{A_{v}}, \widetilde{B_{v}})$
        \STATE $\mathit{partSum} \gets \textsc{ReduceSum}(\mathit{M})$

        \STATE $\mathit{off_{\sum}} \gets \textsc{getOff}_{\sum}(aIdx, bIdx, \mathit{v})$

        \STATE $\textsc{StoreBlock}(\mathit{tmpBuf},\mathit{partSum}, off_{\sum})$
    \ENDFOR
\ENDFOR

\textit{Gather stage}
\STATE $\textsc{WaitAllVectors}()$

\STATE $kParts \gets \mathit{K} \operatorname{DIV} \mathit{kSplit}$
\STATE $\mathit{Acc} \gets \mathbf{0} \in \mathbb{R}^{\mathit{mSplit} \times \mathit{nSplit}}$

\STATE $\mathit{off_C} \gets \textsc{getOff}_{C}(\mathit{v})$

\FOR {$\mathit{i} = 0$ \TO $\mathit{kParts} - 1$}
    \STATE $\mathit{off_{Acc}} \gets \textsc{getOff}_{Acc}(\mathit{v}, i)$
    \STATE $\mathit{partSum} \gets \textsc{LoadBlock}(\mathit{tmpBuf}, off_{Acc})$
    \STATE $\mathit{Acc} \gets \mathit{Acc} + \mathit{partSum}$
\ENDFOR

\text{Stores tensor block in GM by offset}
\STATE $\textsc{StoreBlock}(C, \mathit{Acc}, \mathit{off_C})$
\end{algorithmic}
\end{algorithm}

\subsection{MLA Operator Optimization}

\begin{figure}[htbp!]
    \centerline{\includegraphics[width=0.5\textwidth]{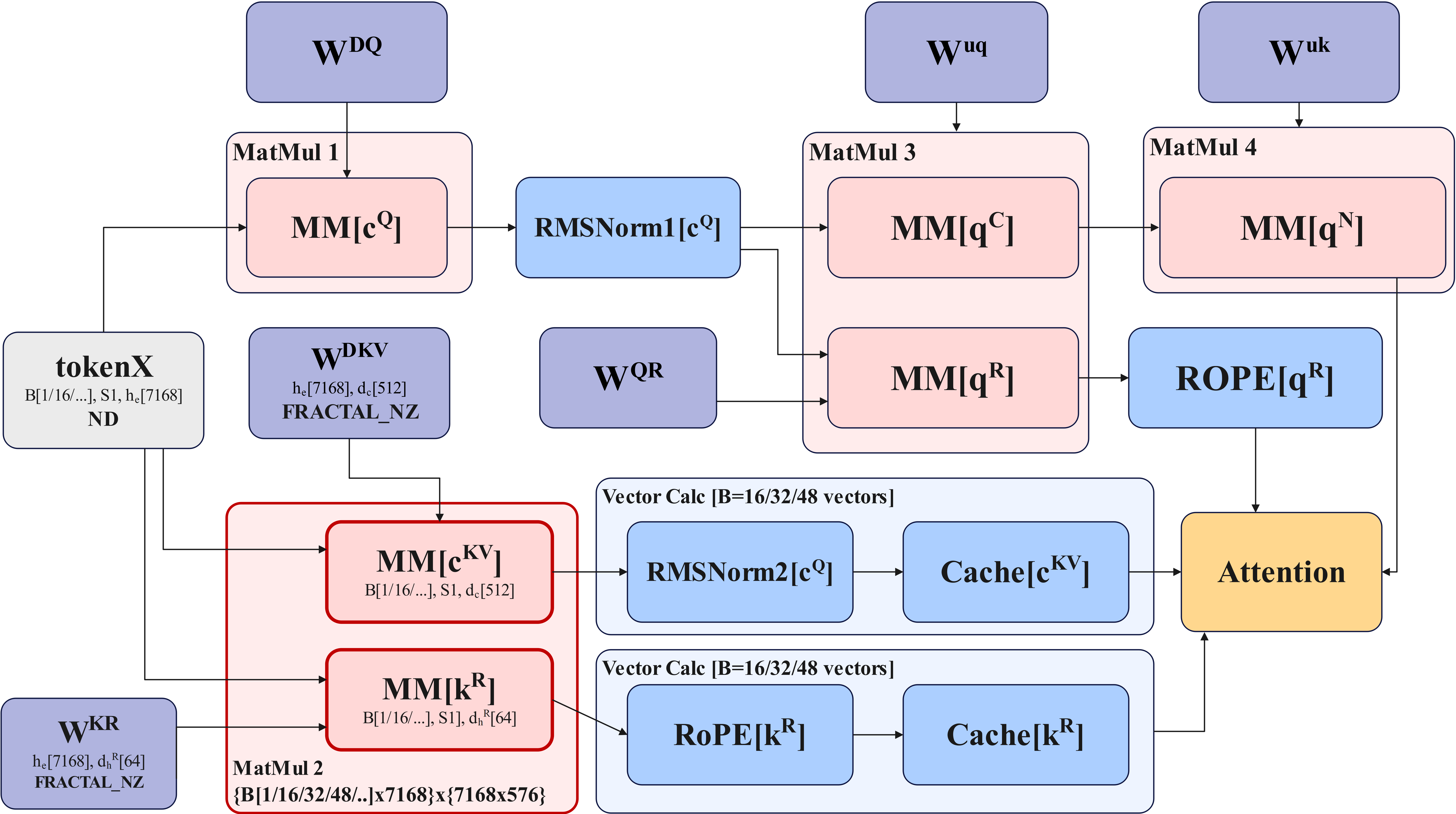}}
    \caption{Baseline MLA operator architecture. The red line highlights two MatMuls that were optimized in Section~\ref{optimization_application}}\label{MLAArch}
\end{figure}

The Multi-Head Latent Attention (MLA) architecture in DeepSeek-V3 is an optimized attention mechanism designed to reduce the KV cache footprint during inference~\cite{liu2024deepseek},
which is critical for efficient long-context processing~\cite{tamkin2021understanding}.
The Fig.~\ref{MLAArch} shows the baseline MLA operator calculation scheme on Ascend AI\@.
Blue color indicates what is calculated on the Vector Unit, and pink color indicates what is calculated on the Cube Unit.

After analyzing the operator's execution on a timing simulator, we determined that the main candidates for optimization are $MM[c^{KV}]$ and $MM[k^{R}]$.
In the original work they correspond to the following equations.
First, it is the multiplication of the input token by the down-projection matrix~\eqref{eq:CV_compression}.

\begin{equation}
\label{eq:CV_compression}
\mathbf{c}_t^{KV} = W^{DKV} \mathbf{h}_t
\end{equation}

Second, it is the multiplication of the input token by the matrix that used to produce the decoupled key that carries Rotary Positional Embedding (RoPE~\cite{su2021roformer})~\eqref{eq:RoPE_KR}.

\begin{equation}
\label{eq:RoPE_KR}
\mathbf{k}_t^R = RoPE(W^{KR} \mathbf{h}_t)
\end{equation}

As an optimization, one can compute these two matrix multiplications as one (\ref{eq:extendedMatMul}), denoting it as \textbf{MatMul 2}, as shown in Fig.~\ref{MLAArch}.

\begin{equation}
\label{eq:extendedMatMul}
\mathbf{MM_{t}} = \left( W^{DKV} | W^{KR} \right) \mathbf{h}_t
\end{equation}

\textbf{Data Independence}: Further calculations on the Cube Unit do not depend on the result of MatMul2. On the contrary, the results go into further calculations on the Vector Unit, such as RMSNorm, RoPE, etc.

\textbf{Idling}: Based on the profiling obtained from the timing simulator, we clearly determined that AIV units are barely used.
In addition, a significant part of the calculations in the Vector Unit depends on the result of MatMul 2.
That is, if the AIV and AIC computations overlap successfully, the calculation of RMS, RoPE and caching normalization can be started immediately after MatMul 2 calculation, without the need for synchronization with the Cube Unit.

\textbf{Narrow Dimension}: The target case of our optimization will be the processing of one token, i.e.\ batch size = 1.

\subsection{Algorithm Specialization}

In this section, we specialize the proposed algorithm in Section~\ref{sec:proposed_algo} for the target case of multiplying a single token by the weight matrices $W^{DKV}$, $W^{KR}$, respectively.
According to the DeepSeek-V3 hyper-parameters~\cite{liu2024deepseek} hidden dimension $d = 7168$, KV compression dimension $d_c = 512$ and for the decoupled queries and key $d_h^R = 64$ was set per-head.

Then:
\[ c_t^{KV} = matmul(W^{DKV}, h_t^T) \]
\[ k_{base} = matmul(W^{KR}, h_t^T) \]
\text{where} $W^{DKV} \in \mathbb{R}^{d_c \times d}, \: W^{KR} \in \mathbb{R}^{d_h^R \times d}, \: h_t^T \in \mathbb{R}^{d \times 1}$.

Thus, equation~\eqref{eq:extendedMatMul} takes the form:
\begin{equation}
\label{eq:MatMul2}
C = T \times W
\end{equation}
\text{where} $T {=} h_t \in \mathbb{R}^{1 \times d}, \: W {=} \left(W^{DKV}|W^{KR}\right)^{T} \in \mathbb{R}^{d \times (d_c + d_h^R)}$.

The specificity of this test case is such that the layout of both weight matrices in memory is a \texttt{FRACTAL\textunderscore{}NZ} tiling (depicted on Fig.~\ref{MatrixDataLayout}), and the data type is \texttt{bfloat16\textunderscore{}t}.
As already mentioned in Section~\ref{subsec:vector_overview}, Vector Unit does not support the \texttt{bfloat16\textunderscore{}t} type for some operations.
In our specialization, \textbf{none of the required vector instructions support this type}, which introduces an extra challenge for applying the optimization.
In the following sections~\ref{subsec:DatпaLayoutIssue},~\ref{subsec:DataTypeIssue} we will look in detail at how we optimally resolved these integration issues into MLA operator workflow without dramatically losing the performance gain..

\subsection{Data Layout Issue}\label{subsec:DataLayoutIssue}

\begin{figure}[htbp!]
    \centerline{\includegraphics[width=0.45\textwidth]{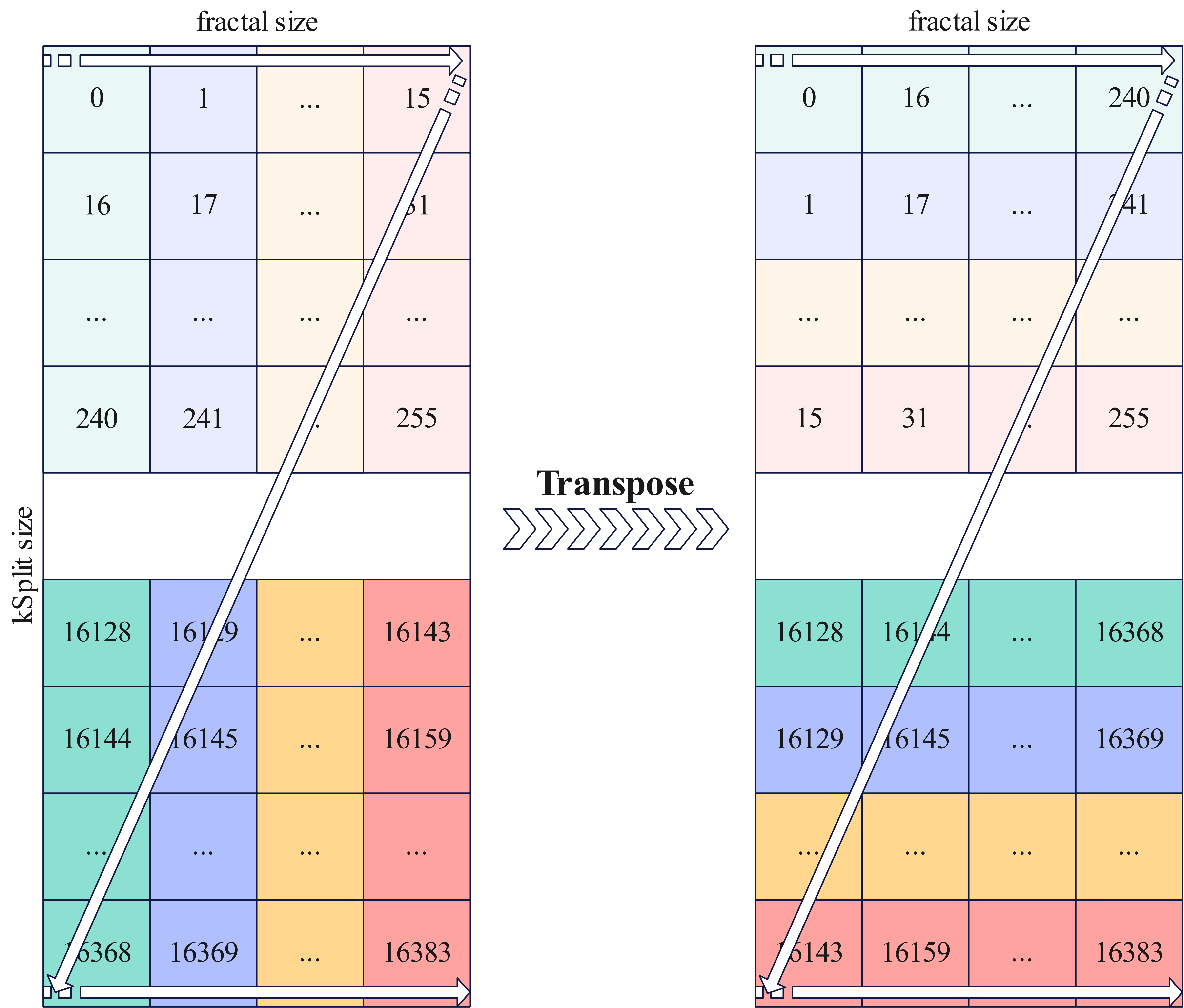}}
    \caption{Fractal Column Transposition Scheme}\label{TransposeScheme}
\end{figure}

For the consistency of the solution, it was necessary to preserve the layout of the matrices in Global Memory, the same as when calculating MatMul 2 on the Cube Unit.
Therefore, the specifics of the \texttt{FRACTAL\textunderscore{}NZ} format, the proposed algorithm (sec.~\ref{sec:proposed_algo}) had to be supplemented with the following transformations:
\begin{itemize}
    \item To get to the proposed row-column multiplication (as in the Algorithm~\ref{alg:generalized_algo}), it is necessary to \textbf{transpose the weight matrix}.
    \item Since the matrix is located in memory in $16\times 16$ fractals, the optimal solution is to transpose it in blocks of the $fractalSize$.
    \item Fortunately, the vector function \texttt{Transpose} from AscendC API can transform $16\times 16$ blocks of two-dimensional tensors of the \texttt{uint16\textunderscore{}t/int16\textunderscore{}t/half} data type.
    \item Unfortunately, the vector function \texttt{Transpose} does not support \texttt{bfloat16\textunderscore{}t}, so the same memory block loaded into UB must first be interpreted as \texttt{half} (Listing~\ref{hack4transpose}).
    \item After this we can transpose the loaded fractal column of size $fractalSize$ $\cdot$ $kSpitSize$ as shown in Fig.~\ref{TransposeScheme} (Listing~\ref{TransoseWeight}).
\end{itemize}

\begin{listing}[!ht]
\begin{lstlisting}[
    language=C++,
    basicstyle=\fontsize{7.5}{7.5}\selectfont\ttfamily,
    aboveskip=0pt,
    belowskip=0pt,
  ]
fp16Kvkr = bf16Kvkr.template ReinterpretCast<half>();
\end{lstlisting}
  \caption{Preliminary cast before weight transposition}
  \label{hack4transpose}
\end{listing}

\begin{listing}[!ht]
\begin{lstlisting}[
    language=C++,
    basicstyle=\fontsize{7.5}{7.5}\selectfont\ttfamily,
    aboveskip=0pt,
    belowskip=0pt,
  ]
for (int i = 0; i < fractalSize; ++i)
    Transpose(fp16KvkrT[256 * i],
              fp16Kvkr[256 * i])
\end{lstlisting}
  \caption{Transpose weight matrix}
  \label{TransoseWeight}
\end{listing}

However, due to the transposition of $16\times 16$ blocks, the column we needed of size $kSplitSize$ turned out to be scattered across the UB-buffer with a block stride equal to the $fractalSize$.
To achieve maximum vector multiplication throughput, this problem was solved using Cast by choosing the repeat stride and block stride parameters appropriately in Section~\ref{subsec:DataTypeIssue}.

\subsection{Data Type Issue}\label{subsec:DataTypeIssue}

\begin{figure}[htbp!]
    \centerline{\includegraphics[width=0.45\textwidth]{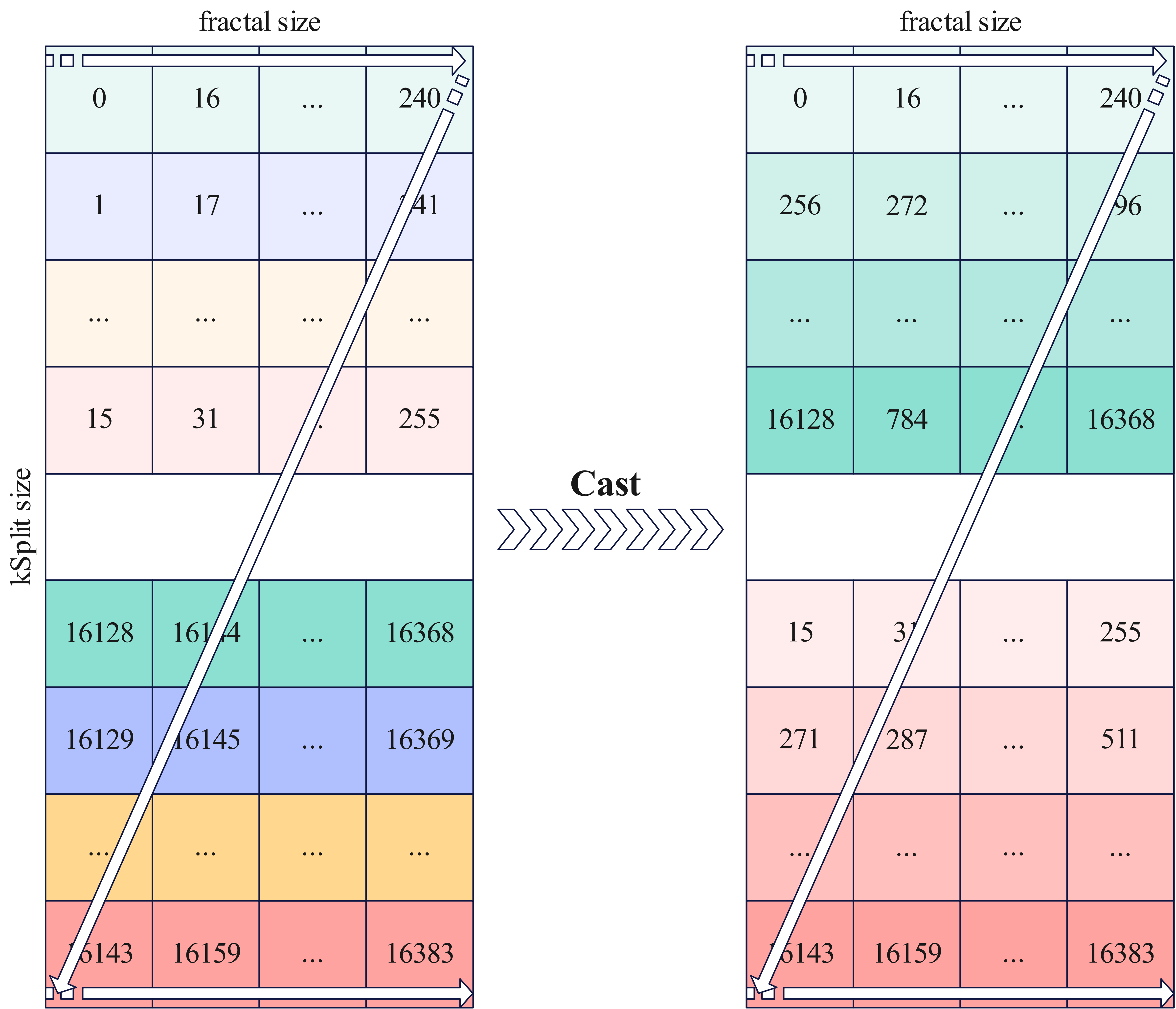}}
    \caption{Fractal Column Cast Scheme. Each \texttt{Cast} call with single repeat collects $\frac{mask}{fractalSize}$ blocks of $fractalSize$.}\label{CastScheme}
\end{figure}

For the consistency of the solution, it was necessary to maintain the calculation accuracy the same as when calculating MatMul 2 on the Cube Unit.
Thus, taking into account the specifics of the data type of the weight matrices and the token row, the proposed algorithm was supplemented with the following calculations:
\begin{itemize}
    \item At the beginning of the \textbf{compute stage}, we convert the data type from \texttt{bfloat16\textunderscore{}t} to \texttt{float} for the
    weight matrix and token row buffers of sizes $fractalSize$ $\cdot$ $kSpitSize$ and $kSpitSize$, respectively.
    Set the rounding mode to \texttt{RoundMode::CAST\textunderscore{}NONE}, because rounding to a type of greater precision.
    \item In addition to type conversion, \texttt{Cast} to the weight matrix block solved another important issue.
    \texttt{Cast} gathers in one repeat $\frac{256}{sizeof(float)}$ (\texttt{mask}) consecutive elements for one column inside the “fractal column” block (see Fig.~\ref{CastScheme}).
    Thus, in $\frac{kSplitSize}{mask}$ (\texttt{opNum}) repeat times \texttt{Cast} not only converts the data to the required format,
    but also collects a whole column of size $kSplitSize$ for subsequent multiplication by the token row (Listing~\ref{CastAll}).
    \item At the end of the \textbf{compute stage}, we store the calculated partial sums to Global Memory in \texttt{float} type (without redundant type conversion).
    \item After accumulating all Vector Unit partial sums at the \textbf{gather stage}, before storing the result to GM, we cast the buffer back to the \texttt{bfloat16\textunderscore{}t} data type.
    Set the rounding mode to \texttt{RoundMode::CAST\textunderscore{}ROUND} so that the accuracy matches that obtained on the Cube Unit.
\end{itemize}

\begin{listing}[!ht]
\begin{lstlisting}[
    language=C++,
    basicstyle=\fontsize{7.5}{7.5}\selectfont\ttfamily,
    aboveskip=0pt,
    belowskip=0pt,
    % columns=fullflexible,
  ]
// dstBlkStride, srcBlkStride, dstRepStride,
//                             srcRepStride
UnaryRepeatParams castParams{1, 16, 8, 64};
int mask = 256 / sizeof(float);
int opNum = kSplitSize / mask;

for (int i = 0; i < fractalSize; ++i)
    Cast(fp32Kvkr[kSplitSize * i]
         bf16KvkrT[fractalSize * i],
         RoundMode::CAST_NONE,
         mask, opNum, castParams);
\end{lstlisting}
  \caption{Cast weight and token tensors from \texttt{bfloat16\textunderscore{}t} to \texttt{float} data type}
  \label{CastAll}
\end{listing}

\subsection{MLA Operator MatMul Algorithm}

\begin{figure}[htbp!]
    \centerline{\includegraphics[width=0.5\textwidth]{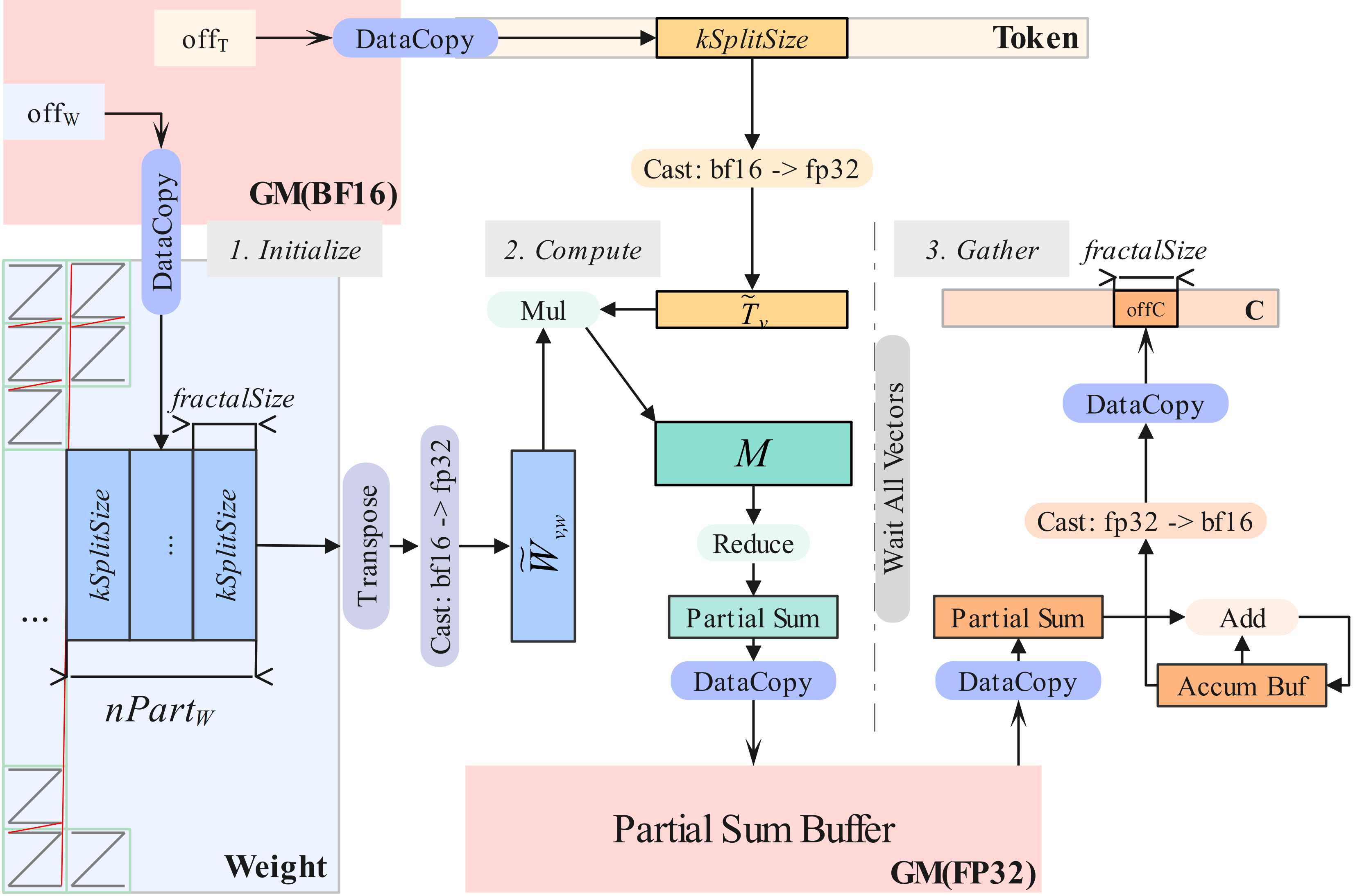}}
    \caption{MatMul 2 (batch size = 1) algorithm scheme on AIV}\label{pic:MMVecAlgorithm}
\end{figure}

We assume that the operator is executed on $P$ physical AI Cores.
The target case of our optimization used exactly 24 physical cores, i.e. $P=24$.
The calculation scheme of MaMul 2 for MLA Operator on Vector Unit is shown in Fig.~\ref{pic:MMVecAlgorithm}.
Due to the specific of the target test, processing a single token in the MLA architecture (Fig.~\ref{MLAArch}), a number of changes were made to the specialized algorithm MatMul calculation (Algorithm~\ref{alg:specialized_matmul_algo}) relative to the generalized solution (Algorithm~\ref{alg:generalized_algo}).
The \textbf{key modifications} to the generalized algorithm are as follows:
\begin{itemize}
    \item Line~\ref{alg:fractalSize}: $fractalSize$ parameter value is taken from the AscendC documentation for \texttt{FRACTAL\textunderscore{}NZ} tiling.
    \item Line~\ref{alg:kSplitSize}: $kSplitSize$ parameter value was chosen empirically so that loading of the weight matrix data via the MTE2 bus at the $w+1$ iteration would overlap with the calculations at the $w$-th iteration.
    \item Line~\ref{alg:VecPool}: parameter $V$ (AIV pool size) is the maximum possible number of vector executors into which the entire weight matrix can be evenly divided into blocks of size $fractalNum \cdot kSplitSize$.
    \item Line~\ref{alg:MatnPartW}: the optimal value of the parameter $V$ allowed us to minimize the number of loop iterations, $nPartW$.
    \item Line~\ref{alg:TokenDataCopy}: $\forall \: v$ loads a token block ($T_v$) from GM to UB only once as an optimization.
    \item Line~\ref{alg:dotProduct}: the multiplication of $\widetilde{T}_v$ by the j-th column of matrix $\widetilde{W}_{v,w}$ and subsequent reduction can be represented as
    a dot product.
    \item Line~\ref{alg:AccInit}$-$\ref{alg:GatherEnd}: since $T$ is a row, then $fractalNum$ number of AIV executors are needed to gather the result matrix $C$, and this is acceptable since $fractalNum < V$
    \item Line~\ref{alg:OutputCast}: accumulation buffer is cast back to the \texttt{bfloat16\textunderscore{}t} type to match the accuracy with Cube Unit and subsequent calculations (RoPE, RMS Norm).
\end{itemize}

\subsection{Evaluation}

Tables~\ref{tab:InstructionSimulation}~and~\ref{tab:TimingSimulation} compare the execution profiles for the baseline and optimized MLA operator on the timing simulator.
The MTE3 instruction count has increased significantly (by 71\%) since AIV does not have an intermediate cache like AIC, but this has not caused any contention\@.
Performance has improved on both the Vector Unit (by 19\%) and the Cube Unit (by 17\%).
Furthermore, despite offloading some matrix calculations to the AIV, the AIC still completes the final computations in the operator.

\begin{figure}[htbp!]
    \centerline{\includegraphics[width=0.5\textwidth]{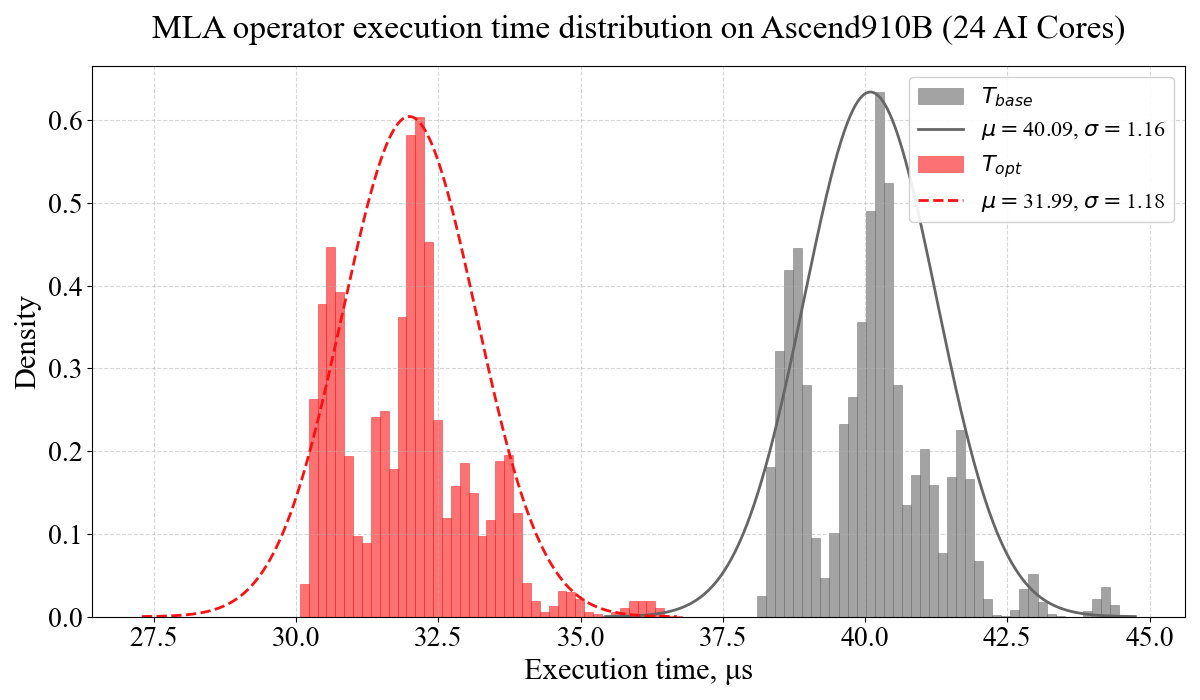}}
    \caption{Statistical distribution of execution times for the baseline and optimized MLA operator on real hardware (Ascend 910B)}\label{pic:StatisticalDistribution}
\end{figure}

The Fig.~\ref{pic:StatisticalDistribution} shows the statistical distribution of execution times for the baseline and optimized MLA operator on real hardware (Ascend 910B).
To analyze the distribution, we performed 100 runs of the operator, with 100 iterations per run.
For each run, we dropped the first iteration to eliminate cold-start effects, such as non-warmed-up caches.
After plotting the execution time distribution graphs for the baseline and optimized versions, we found that the performance gain by the mean value was

\[ G = \left(1 - \frac{\mathop{\mathbb{E}}[T_{opt}]}{\mathop{\mathbb{E}}[T_{base}]}\right) = \left(1 - \frac{31.99}{40.09}\right) \cdot 100 \% \approx 20 \%. \]

\section{Discussion and Future Work}
While our approach demonstrates significant gains, it is not without its trade-offs, which must be considered for application:
\begin{enumerate}
    \item Applying \textbf{additional pressure to the MTE} bus may slow down data transfers to the Cube Unit.
    \item The significant use of Unified Buffer memory (in our case, $\approx 150$ KB per AIV core was consumed) can be critical if the \textbf{UB already contains other data} that the Cube Unit needs and cannot be overwritten.
    \item \textbf{The number of calculations} in the algorithm may increase due to the specific of the data format (redundant \texttt{Cast}) or layout in the GM (redundant \texttt{DataCopy} or \texttt{Transpose}).
\end{enumerate}

\begin{algorithm}[H]
\caption{Specialized MatMul 2 on AIV for MLA operator}
\begin{algorithmic}[1]\label{alg:specialized_matmul_algo}
\REQUIRE $\mathit{T} \in \mathbb{B}^{1 \times K}$, $\mathit{W} \in \mathbb{B}^{K \times N}$
\COMMENT{$\mathbb{B}$: bfloat16} (\ref{eq:MatMul2})

\ENSURE $\mathit{C} \in \mathbb{B}^{1 \times N}$
\COMMENT{$\mathbb{B}$: bfloat16}

\textit{Init stage}

\STATE $\mathit{fractalSize} \gets 16$\label{alg:fractalSize}
\STATE $\mathit{kSplitSize} \gets 1024$\label{alg:kSplitSize}
\STATE $kParts \gets \mathit{K} \operatorname{DIV} \mathit{kSplitSize}$
\STATE $\mathit{fractalNum} \gets N \operatorname{DIV} fractalSize$
\STATE $S \gets fractalNum \cdot kParts$

$\mathit{P}$ \text{-- number of AI cores}

\text{Select the optimal number of AIV executors}
\STATE $V \gets \max\limits_{d \in D_S} \left\{ d \leq 2P \right\}$ \COMMENT{where $D_S$ -- set of divisors $S$}\label{alg:VecPool}

$\mathit{v}$ \text{-- AIV identifier, so} $0 \leq \mathit{v} \leq {V - 1}$

\text{Calculates the number of MatMuls $\forall v$}
\STATE $\mathit{nPart_{W}} \gets S \operatorname{DIV} V$\label{alg:MatnPartW}
\STATE $groupSize \gets \mathit{V} \operatorname{DIV}  kParts$

\text{Calculates token offset according to the $v$}
\STATE $off_{T} \gets  (\mathit{v} \operatorname{DIV} groupSize) \cdot kSplitSize $

\STATE $T_v \gets T[0, \mathit{off_{T}}]$\label{alg:TokenDataCopy}

\textit{Compute stage}

\FOR{$w \gets 0$ \TO $nPart_{W}-1$}
    \STATE $\mathit{fracId} \gets \mathit{v} \cdot nPart_{W} + \mathit{w}$
    \STATE $W_{v,w} \gets W[off_{T},  fracId]$\label{alg:WeightDataCopy}
    \STATE \textbf{Transpose} \text{by fractals:} $W_{v,w}^T \gets W_{v,w}$  \text{ (listing~\ref{TransoseWeight}) }
    \STATE \textbf{Cast} \text{to float:} $\widetilde{T}_v \gets T_v, \widetilde{W}_{v,w} \gets W_{v,w}^T$ \text{ (listing~\ref{CastAll}) }
    \STATE $\mathit{partSum} \gets \left[ \langle \widetilde{T}_v,\widetilde{W}_{v,w}[:,j] \rangle \right]_{j=0}^{fractalSize-1}$\label{alg:dotProduct}
    \STATE $\mathit{tmpBuf}[0, \mathit{fracId}] \gets \mathit{partSum}$
\ENDFOR

\textit{Gather stage}
\STATE $\textsc{WaitAllVectors}()$

\STATE $\mathit{off_{C}} \gets v \cdot fractalSize$
\STATE $\mathit{Acc} \gets \mathbf{0} \in \mathbb{F}^{\mathit{1} \times \mathit{fractalSize}}$\label{alg:AccInit}
\COMMENT{$\mathbb{F}$: float}

\FOR {$i = 0$ \TO $\mathit{kParts} - 1$}
    \STATE $\mathit{partSum} \gets \mathit{tmpBuf}[0, off_{C} + i \cdot N]$
    \STATE $\mathit{Acc} \gets \mathit{Acc} + \mathit{partSum}$
\ENDFOR\label{alg:GatherEnd}

\STATE \textbf{Cast} \text{to bfloat16:} $\mathit{Out} \gets \mathit{Acc}$\label{alg:OutputCast}
\STATE $\mathit{C}[0, \mathit{off_{C}}] \gets \mathit{Out}$
\end{algorithmic}
\end{algorithm}

\begin{table}[htbp]
  \centering
  \caption{Comparison of instruction profiling for the baseline and optimized MLA operator on Ascend AI (timing simulation)}
    \begin{tabular}{cccc}
        \toprule
        \multicolumn{4}{c}{Timing simulation (24 AI Cores)}                                  \\
        \multicolumn{1}{c}{Instructions}  & {Baseline}           & {Optimized} & {$\Delta$}   \\
        \cmidrule(r){1-1}                       \cmidrule(rl){2-4}
          Total                      & 480869              & 565173      &      +84304          \\
          Cube                       & 8442                & 6408        &      -2034           \\
          Vec                        & 1898                & 35594       &      +33696          \\
        Scalar                       & 423763              & 466897      &      +43134          \\
          MTE2                       & 3823                & 3805        &      -18             \\
          MTE3                       & 916                 & 1570        &      +654            \\
        \bottomrule
    \end{tabular}\label{tab:InstructionSimulation}
\end{table}

\begin{table}[htbp]
  \centering
  \caption{Comparison of execution time for the baseline and optimized MLA Operator on Ascend AI (timing simulation)}
    \begin{tabular}{cccc}
        \toprule
        \multicolumn{4}{c}{Timing simulation (24 AI Cores)}                                  \\
        \multicolumn{1}{c}{}  & {Baseline, ms}           & {Optimized, ms} & {Speedup, \%}   \\
        \cmidrule(r){1-1}                       \cmidrule(rl){2-4}
       Total                 & 49.574                 & 40.782    &     17            \\
       AIC                   & 49.573                 & 40.779   &      17             \\
       AIV                   & 49.532                 & 40.105   &      19             \\
        \bottomrule
    \end{tabular}\label{tab:TimingSimulation}
\end{table}

For further work, two hypotheses can be noted.
Let $\mathit{A} \in \mathbb{R}^{M \times K}$, $\mathit{B} \in \mathbb{R}^{K \times N}$ and $d_{\min} = \min(M, N)$.
We assume that optimization may be worthwhile for certain data types and specific layout in memory, if Condition~\ref{condition2} is extended such that:

\begin{definition}[Small Remnants]\label{condition4}
If $d_{\min} \mod 16 = 1$, and the $d_{\min} \neq 1$, then it is worth offloading from AIC to AIV only the narrow part of MatMul that is determined by the remainder, as shown in this paper.
\end{definition}

\begin{definition}[Larger Token]\label{condition5}
If $d_{\min} = 2$.
\end{definition}

\section{Conclusion}
In this work, we have demonstrated that offloading of matrix-vector multiplication from the Cube Unit to the Vector Unit is a effective method for accelerating inference on Ascend AI processors.
The core insight involves offloading of operator's calculations to the idling AIV, allowing them to overlap with subsequent computations on the AIC.\@

The proposed optimization was successfully applied to the MLA operator in DeepSeek-V3.
Despite the challenges related to the specific data layout and type, our approach showed a mean performance gain of 20\% for single-token processing, as confirmed by real hardware measurements.

This research focuses on accelerating inference for server NPUs.\@
We believe our findings contribute a valuable strategy to the CANN Community and this small token multiplication optimization can be extended to another operators with similar computational problems under other NPU chips.

\section{Acknowledgment}
Special thanks to Flegontov Alexander for technical advice and Li Qiduan for administrative support.
We would like to express our gratitude for the reviewers' valuable feedback on our paper.

\bibliographystyle{IEEEtran}
\bibliography{IEEEabrv,references}

\end{document}